\documentclass[journal]{IEEEtran}
\usepackage{amsmath}
\usepackage{amssymb}
\usepackage{graphicx}
\usepackage{booktabs}
\usepackage{hyperref}
\usepackage{geometry}
\usepackage{multirow}
\usepackage{array}
\usepackage{enumitem}
\title{Classical and Quantum Machine Learning for Population-Level Prediction of Heat-Related Physiological Events}
\author{
Sa\'ul Gonz\'alez-Bermejo$^1$,
Tommaso Albrigi$^1$, 
Borja V\'azquez-Morado$^1$,
Urko Regueiro-Ramos$^1$,\\
Daniel Casado-Faul\'\i$^1$,
Sergi C\`onsul-Pacareu$^1$,
Parfait Atchad\'e-Adelomou$^1$\\
(1) Lighthouse Disruptive Innovation Group SL.}
\date{}

\begin{document}

\maketitle

\begin{abstract}

Predicting heat-related physiological events at the population level is challenging due to the complex interactions among climatic, demographic, and socioeconomic factors, as well as the strong sparsity and seasonality of observational data. In this work, we propose a unified predictive framework that integrates heterogeneous environmental and public-health datasets and evaluates two learning paradigms within a common pipeline: classical machine learning and quantum machine learning. The methodology combines data harmonization, temporal aggregation, feature engineering, and dimensionality reduction to construct a weekly county-level population dataset. On this unified representation, we train both a classical regression baseline and a variational quantum model based on parameterized quantum circuits with angle embedding and data re-uploading. Experimental evaluation on datasets from the United States and Catalonia shows that classical models currently achieve higher predictive accuracy, particularly under conditions of strong class imbalance and sparse targets. Nevertheless, the quantum models demonstrate non-trivial learning capability and capture meaningful predictive structure in several scenarios. These results provide an empirical comparison between classical and quantum learning approaches for population-level physiological prediction and establish a methodological foundation for future hybrid health modeling as quantum hardware continues to evolve.

\end{abstract}

\section{Introduction}

Heat-related physiological events and heat-related illnesses constitute an increasingly important public-health concern, particularly under conditions of rising temperatures, demographic aging, urban heat exposure, and heterogeneous social vulnerability \cite{ebi2021heat,siddiqui2025heatreview}. Predicting these events at the population level is relevant for public-health preparedness, territorial risk assessment, and resource planning. However, this task remains challenging because heat-related outcomes are driven by the joint effect of environmental conditions, population structure, and socioeconomic context, while the observable data are often sparse, aggregated at mismatched geographic scales, and strongly seasonal.

In practice, the prediction of heat-related physiological events cannot be reduced to a simple univariate climatic forecasting problem. Instead, it requires the integration of multiple heterogeneous data sources, including meteorological variables, demographic composition, labor-sector structure, and hospitalization-related records. Moreover, the target variable is typically characterized by strong zero inflation, low absolute counts, and temporal concentration during summer periods. These properties make the learning problem statistically difficult and particularly sensitive to modeling assumptions.

Classical machine learning methods offer a natural baseline for this type of task, as they can flexibly model nonlinear relationships from structured tabular data. Boosting-based methods are particularly strong on tabular data and have become standard competitive baselines in many applied prediction settings \cite{ke2017lightgbm}. At the same time, recent advances in variational quantum circuits and hybrid quantum--classical optimization have motivated the exploration of quantum machine learning for regression problems, especially in settings where nonlinear feature interactions and compressed latent representations are relevant \cite{bergholm2022pennylane,schuld2019feature,perezsalinas2020reuploading}. Nevertheless, the practical applicability of quantum machine learning in real public-health prediction tasks remains largely underexplored.

The present work is therefore framed as a \emph{comparative modeling study}. The goal is not to claim a quantum advantage, nor to cast the problem as inherently quantum, but rather to establish a common predictive pipeline in which both classical and quantum models can be trained and evaluated under the same data conditions. This allows us to answer a more grounded research question: \emph{to what extent can current quantum machine learning architectures approximate or complement classical predictive performance in population-level modeling of heat-related physiological events?}

To address this question, we construct a unified weekly county-level population dataset from heterogeneous USA and Catalonia data sources. We then apply a shared preprocessing and dimensionality-reduction pipeline and train two model families: a classical regression baseline and a variational quantum model based on a Quantum Sequential Model (QSM). Their performance is compared across multiple population-level prediction settings.

The main contributions of this work are as follows:
\begin{itemize}
    \item We propose a reproducible population-level data construction pipeline integrating climatological, demographic, economic, and hospitalization-related information from heterogeneous public sources.
    \item We formalize a common preprocessing strategy that supports both classical and quantum learning, including standardization, correlation analysis, and principal-component compression.
    \item We implement and evaluate both classical and quantum regression models under identical experimental conditions.
    \item We provide an empirical comparative analysis that clarifies where classical methods currently dominate and where quantum models already display meaningful learning behavior under NISQ constraints.
\end{itemize}

\section{Related Work}

\subsection{Population-Level Prediction of Heat-Related Health Outcomes}

The prediction of heat-related illness, heat-related hospitalization, and heat-sensitive physiological events has traditionally been studied within environmental epidemiology and public-health risk modeling. A substantial portion of the literature focuses on estimating the effect of temperature extremes, heat waves, and humidity-related stress on morbidity and mortality outcomes through time-series regression and distributed lag non-linear models (DLNMs) \cite{gasparrini2010dlnm,bhaskaran2013timeseries,ebi2021heat}. These approaches have been instrumental in establishing the association between heat exposure and adverse health outcomes, including hospitalization burdens \cite{miyamura2022heat,liao2025hospitalizationburden}.

Although this literature has produced strong explanatory evidence, many studies are not designed as granular predictive systems that integrate heterogeneous environmental, demographic, and socioeconomic data streams. Recent work has started to explore finer-scale geographic variability in heat-related hospitalizations and local vulnerability \cite{hansen2024spatial}, but the predictive integration problem remains open, especially when health outcomes are observed at coarser administrative scales than climate variables.

A second limitation in the existing literature is that many studies focus predominantly on environmental exposure variables while incorporating demographic and socioeconomic factors only in a limited or aggregated way. However, vulnerability to heat is known to depend not only on temperature itself, but also on age structure, labor conditions, urban form, and social inequality \cite{ebi2021heat,clark2024atrisk}. Therefore, realistic population-level predictive frameworks should integrate climate, demography, and socioeconomic structure jointly rather than treating them as separate analytical layers.

\subsection{Classical Machine Learning for Health and Environmental Prediction}

Classical machine learning has become increasingly relevant for structured health and environmental prediction tasks, particularly when complex nonlinear relationships are expected among predictors. Methods such as random forests, gradient boosting, support vector regression, and neural networks have been used in environmental health, hospitalization forecasting, and tabular biomedical prediction problems. Among these methods, boosting-based approaches are particularly effective in structured tabular datasets with mixed feature types and nonlinear dependencies \cite{ke2017lightgbm}. Their strong empirical performance in applied health prediction makes them a natural benchmark for the present work.

However, classical models also face limitations in strongly imbalanced and zero-inflated settings. When the target contains a very high proportion of zeros, standard regression losses may favor conservative predictions, reducing sensitivity to rare but important events. In addition, strong seasonal concentration and low absolute event counts can reduce the effective signal-to-noise ratio, especially after geographic redistribution or aggregation. These issues are directly relevant in the present study.

\subsection{Quantum Machine Learning for Supervised Regression}

Quantum machine learning (QML) has emerged as a promising line of research at the intersection of quantum computing and statistical learning. Within the current NISQ paradigm, the most widely studied approaches are variational quantum circuits, which are parameterized quantum models trained through hybrid classical--quantum optimization loops \cite{bergholm2022pennylane,schuld2019feature,atchade2023fourier}. For supervised regression, a common strategy consists of encoding classical features into a quantum state through an embedding operator, processing the state with a trainable variational circuit, and extracting prediction-relevant information through measurements of observables.

Several studies have shown that such circuits can approximate nonlinear functions and, under some conditions, define expressive quantum feature spaces \cite{schuld2019feature,schuld2020encoding}. Techniques such as angle embedding and data re-uploading are especially relevant because they determine how classical information is mapped to Hilbert space and how expressivity scales under qubit constraints \cite{perezsalinas2020reuploading,jerbi2023beyondkernels}. Yet, despite their conceptual appeal, QML models remain constrained by current hardware limitations, including limited qubit counts, noise accumulation, readout errors, and restrictions on effective circuit depth \cite{mcclean2018barren,wang2021noise,schuld2022advantage}.

\subsection{Variational Circuits, Data Re-Uploading, and NISQ Constraints}

Within the family of variational quantum models, one important distinction concerns the architecture of the ansatz. Some approaches exploit physically motivated inductive biases, such as symmetry-constrained circuits, while others adopt more flexible hardware-efficient or sequential architectures. In the present work, we consider a Quantum Sequential Model (QSM), which separates the circuit into modular layers for data embedding, variational transformation, entanglement, and measurement.

A particularly relevant concept is \emph{data re-uploading}, whereby the classical input is injected multiple times throughout the circuit instead of only once at the beginning \cite{perezsalinas2020reuploading}. This mechanism increases expressivity without necessarily increasing the number of qubits and has been interpreted as the quantum analogue of adding depth in classical neural networks. In practice, data re-uploading is especially valuable in NISQ settings because it provides a way to enrich the representable function class while keeping the circuit architecture relatively compact \cite{perezsalinas2020reuploading,jerbi2023beyondkernels,atchade2023fourier}.

At the same time, the success of variational circuits depends critically on feature scaling, redundancy reduction, and dimension control. Poorly scaled features can saturate rotation angles; redundant features increase gate count without increasing informational content; and large parameter spaces can aggravate training instability and barren-plateau effects \cite{schuld2020encoding,mcclean2018barren,wang2021noise}. Therefore, any fair comparison between classical and quantum models must carefully account for preprocessing and dimensionality-reduction steps that are compatible with both paradigms.

\subsection{Research Gap and Positioning of This Work}

Despite progress in environmental epidemiology, classical machine learning, and quantum learning, there is still a clear gap at their intersection. To the best of our knowledge, there are very few comparative studies that evaluate classical and quantum regression models on a real population-level heat-related health prediction task built from heterogeneous climate, demographic, economic, and hospitalization datasets. Existing benchmarking work in QML repeatedly shows that classical models remain difficult to beat in realistic settings, reinforcing the need for careful problem-specific comparisons rather than broad claims of quantum superiority \cite{schuld2022advantage,ghigliazza2024benchmarking}.

This work addresses that gap by: (i) constructing a common population-level dataset from heterogeneous public and institutional sources; (ii) applying a preprocessing pipeline explicitly designed to support both classical and quantum models; (iii) evaluating both paradigms under identical conditions; and (iv) interpreting the results not as a search for immediate quantum superiority, but as an empirical assessment of current QML capability in a realistic health modeling problem.

\section{Methodology}\label{sec:methodology}

This section describes the methodological framework used to construct the predictive models and to enable a fair comparison between classical and quantum learning approaches. The methodology is structured in three main stages: the definition of the input feature representation, a shared preprocessing pipeline applied to both paradigms, and the implementation of classical and quantum predictive models.

\subsection{Input Feature Representation}

The predictive framework operates on a structured feature vector derived from an integrated dataset combining climatological, demographic, and socioeconomic information. In order to ensure a rigorous comparative evaluation, both the classical and quantum models are trained using exactly the same input representation.

Let the input vector associated with county $c$ and week $w$ be defined as

\begin{equation}
x_{c,w} =
[
x_1, x_2, ..., x_d
],
\end{equation}
where each component corresponds to a feature extracted from the integrated population-level dataset.

The input features can be naturally grouped into three categories reflecting the main drivers of heat-related physiological risk: climatic exposure, demographic structure, and socioeconomic activity.

\subsubsection{Climatic variables}

Climatic features capture the environmental conditions associated with thermal stress and heat exposure. These variables are derived from daily meteorological observations and aggregated to weekly resolution in order to match the temporal granularity of the prediction task.

The climatic variables used in the model include:

\begin{itemize}
\item $t_{max}$: weekly maximum temperature
\item $t_{mean}$: weekly mean temperature
\item $t_{min}$: weekly minimum temperature
\item $vp$: vapor pressure
\item $vp_{sat}$: saturated vapor pressure
\item $rh$: relative humidity
\item $heatwave\_indicator$: binary indicator of heatwave occurrence
\item $days\_p95$: number of days exceeding the 95th temperature percentile
\end{itemize}

Together, these variables characterize both baseline thermal conditions and extreme heat exposure events.

\subsubsection{Demographic variables}

Demographic variables describe the population structure of each geographic unit and capture potential differences in vulnerability across age groups.

The demographic variables included in the model are:

\begin{itemize}
\item $pop\_total$: total population
\item $pop\_male$: male population
\item $pop\_female$: female population
\item $pop\_age\_0\_17$: population aged 0–17
\item $pop\_age\_18\_64$: population aged 18–64
\item $pop\_age\_65\_plus$: population aged 65 and above
\end{itemize}

To mitigate scale effects and avoid multicollinearity associated with raw population counts, these variables are converted into proportional ratios relative to the total population of the geographic unit.

\subsubsection{Economic structure variables}

Socioeconomic structure is represented through aggregated employment sector indicators derived from census labor statistics. These variables provide information about the local economic composition and potential occupational exposure to heat.

The following sector categories are included:

\begin{itemize}
\item $sector\_agriculture$
\item $sector\_construction$
\item $sector\_industry$
\item $sector\_services$
\end{itemize}

These variables capture structural characteristics of local labor activity that may influence exposure to extreme heat conditions.

\subsubsection{Gaussian seasonal feature variables}

Surveillance data on heat-related illnesses are typically collected and reported on an annual basis for administrative jurisdictions within each country. In the United States, these data are aggregated at the state level, while in Spain they are reported at the provincial level. As a consequence, higher-resolution information—both in terms of temporal granularity (e.g., weekly incidence) and finer geographic units—is generally unavailable. This limitation primarily arises from patient data privacy and confidentiality requirements, which restrict the dissemination of detailed health records that could potentially allow individual identification.

We model the temporal evolution of heat-related illnesses (HRIs) using a structured latent intensity formulation that separates climatological seasonality, county-specific vulnerability, and transient heatwave shocks. For a county (c) in year (y), the expected HRI intensity at time (t) is defined as:
\begin{equation}
    w_{c,y}(t) = g_{\text{season}}(t), m_{c,y}(t) + \sum_{i \in HW_{c,y}} g_{hw}(t - t_i)
\end{equation}
where ($g_{\text{season}}(t)$) represents the shared seasonal climatology of heat-related risk, ($m_{c,y}(t)$) captures spatially heterogeneous vulnerability factors, and the summation term models the contribution of discrete heatwave events occurring at times ($t_i$). The seasonal component ($g_{\text{season}}(t)$) is modeled as a Gaussian kernel:
\begin{equation}
    g_{\text{season}}(t) = \exp\left(-\frac{(t-\mu)^2}{2\sigma^2}\right),
\end{equation}
where ($\mu$) corresponds to the climatological peak of summer (typically mid-July) and ($\sigma$) controls the temporal spread of seasonal risk (with 92\% of these visits occurring during May–September) \cite{seasonal_dist}. This formulation reflects the empirical observation that heat-related morbidity follows a smooth, unimodal annual cycle driven by large-scale temperature dynamics. The multiplicative term ($m_{c,y}(t)$) represents county-level exposure and vulnerability and is parameterized as a function of demographic, socioeconomic, and environmental covariates ($X_{c,y,t}$), typically through a log-linear mapping ($m_{c,y}(t)=\exp(\beta^\top X_{c,y,t})$), allowing the seasonal baseline to be locally amplified or attenuated. Extreme heat events introduce short-term deviations from the climatological baseline, which we model through a response kernel:

\begin{equation}
\label{eq:seasonal_gaussian}
    g_{hw}(\Delta t) = \alpha \exp(-\lambda \Delta t),\quad\mathbf{1}*{{\Delta t \ge 0}},
\end{equation}

representing an immediate increase in HRI incidence followed by an exponential decay over subsequent days. Observed counts (Y*{c,y}(t)) are then assumed to arise from a stochastic observation model centered on this latent intensity, e.g.

\begin{equation}
Y_{c,y}(t) \sim \text{NegBin}\left(w_{c,y}(t), \theta\right),
\end{equation}

which accounts for over-dispersion commonly observed in epidemiological count data. This generative formulation produces a realistic distribution for HRIs by embedding known physical and epidemiological mechanisms—seasonal climatology, spatial vulnerability, and episodic heat extremes—while allowing stochastic variability around the latent intensity. Consequently, the resulting target variable provides a principled and physically interpretable training signal for downstream predictive models.

At the feature level, seasonal variables were incorporated to explicitly represent the underlying meteorological cycle:

\begin{itemize}
\item $season\_gaussian(w)$: Seasonal gaussian distribution explained in Eq. \ref{eq:seasonal_gaussian}
\item $hw\_kernel(w)$: Presence of heatwave close to week $w$
\end{itemize}

By providing the model with structured information about seasonal patterns, the learning algorithm can focus on modeling deviations around the expected climatological behavior rather than attempting to infer the seasonal structure directly from the data. This approach reduces the risk of overfitting and allows the model to capture residual variability associated with short-term anomalies or extreme events more effectively.

\subsubsection{Final feature vector}

The final input representation can therefore be expressed as

\begin{equation}
x_{c,w} =
\begin{bmatrix}
x^{(climate)}_{c,w} \\
x^{(demographic)}_{c,w} \\
x^{(economic)}_{c,w} \\
x^{(seasonal)}_{c,w}
\end{bmatrix}.
\end{equation}

This structured representation provides a compact yet expressive description of the environmental and population context associated with each county-week observation.

\subsection{Common Preprocessing Pipeline}

A central design principle of this study is that both classical and quantum models must operate on an identical feature space in order to enable a meaningful comparison. Consequently, all preprocessing steps are applied prior to model training and are shared by both paradigms.

Let

\begin{equation}
X = \{x_1,\dots,x_N\}, \qquad x_i \in \mathbb{R}^{d},
\end{equation}
denote the dataset after feature construction and filtering.

First, all features are standardized using z-score normalization:

\begin{equation}
x_j' = \frac{x_j - \mu_j}{\sigma_j},
\end{equation}
where $\mu_j$ and $\sigma_j$ denote the empirical mean and standard deviation of feature $j$. Standardization is required to ensure numerical stability during model training and is particularly important in the quantum setting, where features are encoded as rotation angles in parameterized quantum gates \cite{schuld2020encoding,bergholm2022pennylane,consulpacareu2023quantummachinelearninghyperparameter}.

Next, a correlation analysis is performed to identify redundant variables and reduce multicollinearity within the feature space.

Finally, dimensionality reduction is applied using Principal Component Analysis (PCA) \cite{jolliffe2002pca}. PCA projects the standardized feature space into a lower-dimensional representation:

\begin{equation}
\tilde{x} = W^\top x',
\end{equation}
where $W$ denotes the matrix of principal component directions. The number of retained components $k$ is selected so that approximately $98\%$ of the total variance of the original feature space is preserved:

\begin{equation}
\sum_{i=1}^{k} \lambda_i \approx 0.98 \sum_{i=1}^{d} \lambda_i.
\end{equation}

This dimensionality reduction step is particularly relevant for the quantum learning pipeline, where current NISQ hardware imposes strict limitations on the number of qubits and circuit depth \cite{mcclean2018barren,wang2021noise}. The resulting compressed representation

\begin{equation}
\tilde{x}_{c,w} \in \mathbb{R}^{k},
\end{equation}
is used as input for both classical and quantum predictive models.

\subsection{Classical Predictive Model}

The classical model serves as the reference benchmark in the comparative evaluation. Given an input vector ${x}_i' \in \mathbb{R}^{d}$, the model produces a prediction

\begin{equation}
\hat{y}_i^{(C)} = f_{\theta_C}({x}_i'),
\end{equation}
where $f_{\theta_C}$ denotes a regression function parameterized by $\theta_C$.

Model training is performed by minimizing the mean squared error loss

\begin{equation}
\mathcal{L}_C = \frac{1}{N}\sum_{i=1}^{N}\left(\hat{y}_i^{(C)} - y_i\right)^2.
\end{equation}

In the present implementation, the classical baseline is based on the Light Gradient Boosting Machine (LightGBM), a tree-based gradient boosting framework designed for high efficiency and strong predictive performance on structured tabular datasets \cite{ke2017lightgbm}. Gradient boosting methods are particularly well-suited to heterogeneous feature spaces and nonlinear dependencies, making them a strong baseline for population-level prediction tasks.

\subsection{Quantum Predictive Model}

The quantum learning architecture is implemented using a variational Quantum Sequential Model (QSM), which consists of an ordered composition of parameterized quantum layers.

Formally, the quantum model is defined as

\begin{equation}
U(\tilde{x},\theta_Q) =
L_K(\theta_Q^{(K)}) \circ \cdots \circ L_2(\theta_Q^{(2)}) \circ L_1(\tilde{x},\theta_Q^{(1)}),
\end{equation}
where each layer $L_k$ represents a quantum transformation acting on the qubit register.

The first stage of the circuit performs classical-to-quantum feature encoding through angle embedding:

\begin{equation}
U_{\mathrm{enc}}(\tilde{x}) = \prod_{i=1}^{k} R_Y(\tilde{x}_i),
\end{equation}
where $R_Y(\tilde{x}_i)$ denotes a rotation around the $Y$ axis. This encoding maps the PCA-compressed classical feature vector into the quantum Hilbert space \cite{schuld2020encoding,bergholm2022pennylane}.

After the encoding stage, the circuit applies trainable variational layers

\begin{equation}
V(\theta) =
\prod_q
R_X(\theta_{q,1})
R_Y(\theta_{q,2})
R_Z(\theta_{q,3})
\cdot U_{\mathrm{ent}},
\end{equation}
where $U_{\mathrm{ent}}$ denotes an entangling operation typically implemented through controlled-NOT (CNOT) gates.

To increase the expressive capacity of the model without increasing the number of qubits, the architecture employs a data re-uploading strategy. In this setting, the classical features are injected multiple times within the circuit:

\begin{equation}
U(\tilde{x},\theta_Q) =
\prod_{\ell=1}^{L}
V(\theta_Q^{(\ell)})
U_{\mathrm{enc}}(\tilde{x}).
\end{equation}

Data re-uploading has been shown to significantly improve the representational power of variational quantum circuits within the constraints of NISQ hardware \cite{perezsalinas2020reuploading,jerbi2023beyondkernels}.

Predictions are obtained by measuring expectation values of quantum observables

\begin{equation}
z_j(\tilde{x}) =
\langle 0 | U^\dagger(\tilde{x},\theta_Q) O_j U(\tilde{x},\theta_Q) |0\rangle,
\end{equation}

which are then combined through a classical regression mapping

\begin{equation}
\hat{y}_i^{(Q)} = g(z_1,\dots,z_m).
\end{equation}

The quantum model is trained using the same mean squared error loss used in the classical baseline:

\begin{equation}
\mathcal{L}_Q =
\frac{1}{N}\sum_{i=1}^{N}
\left(\hat{y}_i^{(Q)} - y_i\right)^2.
\end{equation}

\section{Experimental Setup and Results}

This section describes the experimental design used to evaluate the predictive performance of the proposed framework and presents a comparative analysis between classical and quantum learning models.

\subsection{Experimental Setup}

The experimental evaluation follows a unified pipeline designed to ensure a consistent comparison between classical and quantum models. Both approaches share the same feature representation and preprocessing steps described in Section~\ref{sec:methodology}.

Two population-level datasets are considered in the study:

\begin{itemize}
\item \textbf{USA population dataset}: county-week observations constructed from CDC health data (Survey USA), Daymet climate data, and U.S. Census demographic information.
\item \textbf{Catalonia population dataset}: A comarca-level, weekly dataset has been constructed by integrating climatological data from Meteocat, demographic statistics from Idescat, and health data derived from the Hospital Morbidity Survey (EMH). The EMH is conducted by the Spanish National Statistics Institute, in collaboration with Eustat in the Autonomous Community of the Basque Country, and with CatSalut and thon the United States dataset, which provides substantially more Spanish National Statistics Institute (INE).
\end{itemize}

A key aspect of the experimental design is the separation between the dataset used for model training and the dataset used for prediction and inference. The predictive models are trained using the United States dataset, which provides a substantially larger number of observations and greater geographic coverage. This larger dataset enables the models to learn stable relationships between climatic exposure, population structure, and heat-related physiological events.

In order to reduce the distributional mismatch between the training and inference domains, the training data are restricted to U.S. states whose climatic characteristics are comparable to those observed in Catalonia. This filtering step ensures that the environmental conditions represented in the training dataset are statistically consistent with the climatic regime of the target region.

Formally, let

\begin{equation}
D_{US}^{train}
\end{equation}
denote the subset of the U.S. dataset used for model training, and

\begin{equation}
D_{CAT}^{test}
\end{equation}
the Catalonia dataset used for prediction and inference. The training process can therefore be expressed as

\begin{equation}
f_{\theta} = \mathcal{A}(D_{US}^{train}),
\end{equation}
where $\mathcal{A}$ denotes the learning algorithm. Model predictions are subsequently obtained for the Catalonia dataset as

\begin{equation}
\hat{y}_{CAT} = f_{\theta}(D_{CAT}^{test}).
\end{equation}

The prediction target corresponds to the weekly aggregated number of heat-related physiological events derived from the population-level dataset.

The experimental workflow consists of the following stages:

\begin{enumerate}
\item Construction of the county-week population dataset
\item Feature preprocessing (standardization and correlation filtering)
\item Dimensionality reduction using PCA (as a preprocessing step for quantum computing)
\item Training of classical and quantum predictive models on the U.S. dataset
\item Prediction and inference on the Catalonia dataset
\item Evaluation of predictive performance
\end{enumerate}

The quantum models are implemented using the UniQuE experimentation platform \cite{casado2025unique}. UniQuE is a hybrid research framework designed to support rapid prototyping of quantum and hybrid quantum--classical machine learning models within controlled experimental pipelines \cite{casadoapplying}.

\subsection{Evaluation Metrics}

Model performance is evaluated using standard regression metrics commonly used in machine learning:

\begin{itemize}

\item \text{Mean Absolute Error (MAE)}: measures the average absolute difference between predictions and observed values.
\item \text{Coefficient of determination ($R^2$)}: measures the proportion of variance in the target variable explained by the model.
\end{itemize}

Formally, the coefficient of determination is defined as

\begin{equation}
R^2 = 1 - \frac{\sum_i (y_i - \hat{y}_i)^2}{\sum_i (y_i - \bar{y})^2},
\end{equation}

while the mean absolute error is defined as

\begin{equation}
MAE = \frac{1}{N} \sum_{i=1}^{N} |y_i - \hat{y}_i|.
\end{equation}

In addition to these metrics, residual distributions and tolerance-based error analyses are used to qualitatively assess prediction stability and error concentration.

\subsection{Results: Catalonia Population Model}

The first experimental scenario evaluates the models on the Catalonia population dataset.

Under the experimental conditions considered, both models face structural difficulties due to the sparsity and irregularity of the target variable. Nevertheless, the classical model exhibits a more stable behavior and a tighter concentration of errors.

\begin{figure*}[h!]
    \centering
    \includegraphics[width=0.4\linewidth]{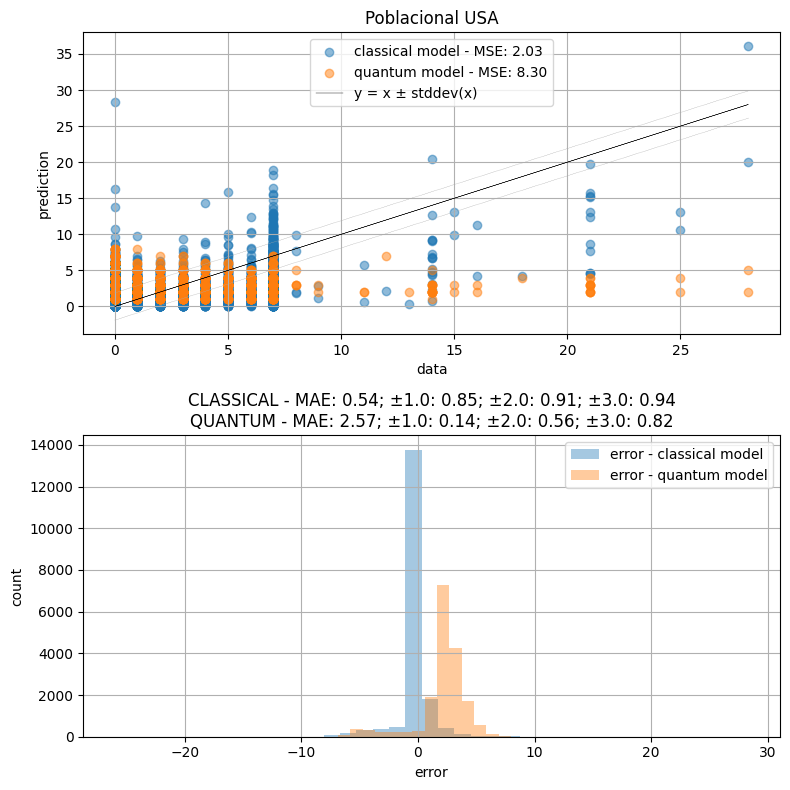}
    \includegraphics[width=0.4\linewidth]{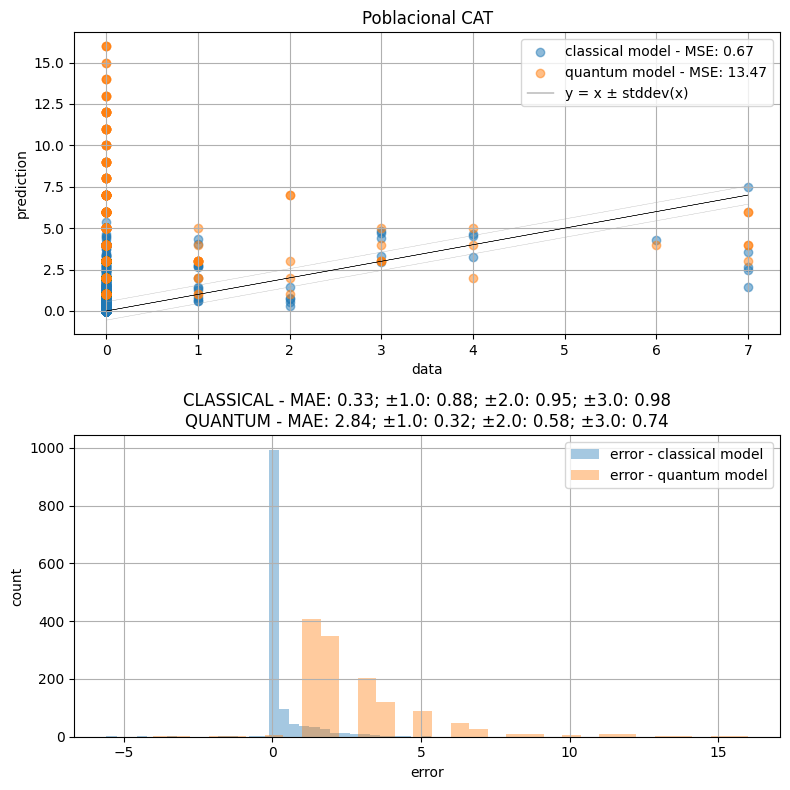}
    \caption{ Comparison between classical (blue) and quantum (orange) models across multiple regression tasks.  Top panels show predicted versus observed values, while bottom panels display the corresponding error distributions. The diagonal line indicates perfect prediction ($y=x$). Classical models outperform quantum models in the population-level tasks (CAT and USA). 
    }
    \label{fig:poblacional-cat}
\end{figure*}

The classical model achieves

\begin{equation}
MAE = 0.33,
\end{equation}

while the quantum model obtains

\begin{equation}
MAE = 2.84.
\end{equation}

Both models yield negative $R^2$ values, indicating that the predictive task remains challenging under the available feature representation. However, the classical model maintains significantly lower prediction errors and a more concentrated residual distribution.

These results suggest that the Catalonia dataset presents a highly complex regression landscape, likely influenced by low event counts and high noise levels in the reconstructed weekly signal.

\subsection{Results: USA Population Model}

The second experimental scenario evaluates the models on the USA population dataset, which contains a larger number of counties and observations.

In this case, the classical model achieves a moderate explanatory capacity, while the quantum model shows a weaker predictive fit.

The classical model reaches

\begin{equation}
MAE = 0.54,
\end{equation}

while the quantum model obtains

\begin{equation}
MAE = 2.57.
\end{equation}

The results indicate a clear advantage of the classical model in terms of both explanatory power and prediction stability. The quantum model exhibits a higher dispersion of errors and a weaker alignment with the target variable.

Nevertheless, it is important to note that the quantum model still captures a non-trivial functional relationship with the target variable, suggesting that the learning process is not purely random but limited by current architectural and hardware constraints.

\subsection{Discussion of Results}

The observed performance differences must be interpreted within the technological constraints of current quantum machine learning implementations.

Variational quantum circuits operate under significant limitations imposed by the Noisy Intermediate-Scale Quantum (NISQ) regime, including restricted circuit depth, limited qubit counts, and noise sensitivity. These constraints directly affect the expressive capacity of the quantum models.

Furthermore, the prediction target used in this study exhibits strong sparsity and seasonal concentration, which significantly increases the difficulty of the regression task. Population-level heat-related events are rare and unevenly distributed across time and geography, making them challenging to model even with classical machine learning approaches.

Despite these limitations, the quantum models demonstrate a meaningful learning signal and provide an important experimental benchmark. The results therefore should not be interpreted as a failure of quantum learning, but rather as an empirical characterization of its current capabilities when applied to real-world population health prediction tasks.

As quantum hardware continues to evolve, hybrid classical–quantum pipelines such as the one proposed in this work may become increasingly relevant for complex predictive modeling problems involving heterogeneous data sources.

\section{Code Availability}

The code used to reproduce the experiments presented in this study is publicly available on GitHub:

\begin{center}
\url{https://github.com/Lighthouse-DIG/PredictHealthToilet-PHT}
\end{center}

Classical models were implemented using Microsoft Azure Machine Learning, while the quantum machine learning models were implemented within the UniQuE experimentation framework and evaluated using quantum circuit simulators.

\section*{Acknowledgment}

This research was partially supported by the Subvencions RETECH program. The authors thank the National Statistics Institute (INE) in collaboration with Eustat – the Basque Statistics Institute in the Autonomous Community of the Basque Country, and in collaboration with CatSalut and the Department of Health of the Government of Catalonia in Catalonia, for conducting and providing the Hospital Morbidity Survey (EMH).

\newpage
\bibliographystyle{unsrt}
\bibliography{population_model_refs}

@book{jolliffe2002pca,
  author    = {Ian T. Jolliffe},
  title     = {Principal Component Analysis},
  edition   = {2nd},
  publisher = {Springer},
  address   = {New York},
  year      = {2002}
}

@inproceedings{ke2017lightgbm,
  author    = {Guolin Ke and Qi Meng and Thomas Finley and Taifeng Wang and Wei Chen and Weidong Ma and Qiwei Ye and Tie{-}Yan Liu},
  title     = {{LightGBM}: A Highly Efficient Gradient Boosting Decision Tree},
  booktitle = {Advances in Neural Information Processing Systems 30 (NeurIPS 2017)},
  year      = {2017}
}

@article{gasparrini2010dlnm,
  author  = {Antonio Gasparrini and Ben Armstrong and Michael G. Kenward},
  title   = {Distributed lag non-linear models},
  journal = {Statistics in Medicine},
  volume  = {29},
  number  = {21},
  pages   = {2224--2234},
  year    = {2010}
}

@article{bhaskaran2013timeseries,
  author  = {Krishnan Bhaskaran and Antonio Gasparrini and Shakoor Hajat and Liam Smeeth and Ben Armstrong},
  title   = {Time series regression studies in environmental epidemiology},
  journal = {International Journal of Epidemiology},
  volume  = {42},
  number  = {4},
  pages   = {1187--1195},
  year    = {2013}
}

@article{ebi2021heat,
  author  = {Kristie L. Ebi and Shakoor Hajat and Jason M. Hess and others},
  title   = {Hot weather and heat extremes: health risks},
  journal = {The Lancet},
  volume  = {398},
  number  = {10301},
  pages   = {698--708},
  year    = {2021}
}

@article{siddiqui2025heatreview,
  author  = {Syed A. Siddiqui and others},
  title   = {A systematic review and meta-analysis of the impact of heat exposure on morbidity and mortality associated with non-communicable diseases in low- and middle-income countries},
  journal = {Environmental Research},
  volume  = {278},
  pages   = {121319},
  year    = {2025}
}

@article{miyamura2022heat,
  author  = {Kei Miyamura and others},
  title   = {Association between heat exposure and hospitalization for hyperglycemic emergencies in Japan: a nationwide study},
  journal = {Environmental Health Perspectives},
  volume  = {130},
  number  = {9},
  pages   = {097008},
  year    = {2022}
}

@article{liao2025hospitalizationburden,
  author  = {Shanshan Liao and others},
  title   = {Temperature-related hospitalization burden under climate change},
  journal = {Nature},
  year    = {2025}
}

@article{hansen2024spatial,
  author  = {K. Hansen and others},
  title   = {The spatial distribution of heat related hospitalizations and local vulnerability across a wide area},
  journal = {Environmental Research},
  year    = {2024}
}

@article{clark2024atrisk,
  author  = {A. Clark and others},
  title   = {Identifying groups at-risk to extreme heat: Intersections of social vulnerability and heat-related health impacts},
  journal = {Environmental Research},
  year    = {2024}
}

@article{bergholm2022pennylane,
  author  = {Ville Bergholm and Josh Izaac and Maria Schuld and Christian Gogolin and Shahnawaz Ahmed and others},
  title   = {PennyLane: Automatic differentiation of hybrid quantum-classical computations},
  journal = {arXiv preprint arXiv:1811.04968},
  year    = {2022}
}

@article{schuld2019feature,
  author  = {Maria Schuld and Nathan Killoran},
  title   = {Quantum machine learning in feature Hilbert spaces},
  journal = {Physical Review Letters},
  volume  = {122},
  number  = {4},
  pages   = {040504},
  year    = {2019}
}

@article{schuld2020encoding,
  author  = {Maria Schuld and Ryan Sweke and Johannes Jakob Meyer},
  title   = {The effect of data encoding on the expressive power of variational quantum machine learning models},
  journal = {Physical Review A},
  volume  = {103},
  number  = {3},
  pages   = {032430},
  year    = {2021}
}

@article{perezsalinas2020reuploading,
  author  = {Adri{\'a}n P{\'e}rez-Salinas and Alba Cervera-Lierta and Elies Gil-Fuster and Jos{\'e} I. Latorre},
  title   = {Data re-uploading for a universal quantum classifier},
  journal = {Quantum},
  volume  = {4},
  pages   = {226},
  year    = {2020}
}

@article{jerbi2023beyondkernels,
  author  = {Sofiene Jerbi and Casper Gyurik and Simon C. Benjamin and Vedran Dunjko and Maria Schuld},
  title   = {Quantum machine learning beyond kernel methods},
  journal = {Nature Communications},
  volume  = {14},
  pages   = {517},
  year    = {2023}
}

@article{mcclean2018barren,
  author  = {Jarrod R. McClean and Sergio Boixo and Vadim N. Smelyanskiy and Ryan Babbush and Hartmut Neven},
  title   = {Barren plateaus in quantum neural network training landscapes},
  journal = {Nature Communications},
  volume  = {9},
  pages   = {4812},
  year    = {2018}
}

@article{wang2021noise,
  author  = {Samson Wang and Enrico Fontana and M. Cerezo and Kunal Sharma and Andrew Sornborger and Patrick J. Coles and Lakshminarayan Subramanian},
  title   = {Noise-induced barren plateaus in variational quantum algorithms},
  journal = {Nature Communications},
  volume  = {12},
  pages   = {6961},
  year    = {2021}
}

@article{schuld2022advantage,
  author  = {Maria Schuld and Nathan Killoran and others},
  title   = {Is quantum advantage the right goal for quantum machine learning?},
  journal = {PRX Quantum},
  volume  = {3},
  number  = {3},
  pages   = {030101},
  year    = {2022}
}

@article{ghigliazza2024benchmarking,
  author  = {Rocco Ghigliazza and Giuseppe Carleo and others},
  title   = {Better than classical? The subtle art of benchmarking quantum machine learning models},
  journal = {arXiv preprint arXiv:2403.07059},
  year    = {2024}
}

@article{casado2025unique,
  title={UniQuE: A General-Purpose Platform for Benchmarking Classical and Quantum Machine Learning Algorithms},
  author={Casado-Faulí, Daniel and González-Bermejo, Saúl and Rovira, Marc and Mei, Alba and Coronas Sala, Laia and Cònsul-Pacareu, Sergi and Vilella, Esteve and Albó-Canals, Jordi and Atchade-Adelomou, Parfait and Vilasis-Cardona, Xavier and Golobardes Ribe, Elisabet},
  journal={TechRxiv},
  year={2025}
}

@article{atchade2023fourier,
  title={Fourier series weight in quantum machine learning},
  author={Atchade-Adelomou, Parfait and Larson, Kent},
  journal={arXiv preprint arXiv:2302.00105},
  year={2023}
}

@misc{consulpacareu2023quantummachinelearninghyperparameter,
      title={Quantum Machine Learning hyperparameter search}, 
      author={S. Consul-Pacareu and R. Montaño and Kevin Rodriguez-Fernandez and Àlex Corretgé and Esteve Vilella-Moreno and Daniel Casado-Faulí and Parfait Atchade-Adelomou},
      year={2023},
      eprint={2302.10298},
      archivePrefix={arXiv},
      primaryClass={cs.LG},
      url={https://arxiv.org/abs/2302.10298}, 
}

@article{casadoapplying,
  title={Applying the UniQuE platform to prototype a solution for the WeShareCare Use Case},
  year={2025},
  author={Casado-Faul{\'\i}, Daniel and Gonzalez-Bermejo, Saul and Rovira, Marc and Mei, Alba and Sala, Laia Coronas and Consul-Pacareu, Sergi and Vilella, Esteve and Albo-Canals, Jordi and Atchade-Adelomou, Aarfait},
  journal={Authorea Preprints},
  publisher={Authorea}
}

@article{seasonal_dist,
  author  = {Vaidyanathan, A. and Gates, A. and Brown, C. and Prezzato, E. and Bernstein, A.},
  title   = {Heat-Related Emergency Department Visits},
  journal = {CDC Morbidity and Mortality Weekly Report},
  pages   = {324--329},
  doi   = {http://dx.doi.org/10.15585/mmwr.mm7315a1},
  year    = {2024}
}

\end{document}